\documentclass[aps,prl,reprint,showpacs,floatfix,superscriptaddress,citeautoscript]{revtex4-1}
\usepackage{graphicx}
\usepackage{bm,amsmath,amssymb,mathrsfs,dcolumn}
\usepackage{epsfig}
\usepackage{hyperref}
\usepackage[usenames]{color} 
\hypersetup{pdfborder=0 0 0,colorlinks=true,citecolor=blue,linkcolor=blue}
\usepackage[final]{pdfpages}

\begin{document}
\title{Gilbert damping in noncollinear ferromagnets}
\author{Zhe Yuan}
\altaffiliation{Present address: Institut f{\"u}r Physik, Johannes Gutenberg--Universit{\"a}t Mainz, Staudingerweg 7, 55128 Mainz, Germany}
\email{zyuan@uni-mainz.de}
\affiliation{Faculty of Science and Technology and MESA$^+$ Institute for Nanotechnology, University of Twente, P.O. Box 217, 7500 AE Enschede, The Netherlands}
\author{Kjetil M. D. Hals}
\affiliation{Department of Physics, Norwegian University of Science and Technology, NO-7491 Trondheim, Norway}
\affiliation{Niels Bohr International Academy and the Center for Quantum Devices, Niels Bohr Institute, University of Copenhagen, 2100 Copenhagen, Denmark }
\author{Yi Liu}
\affiliation{Faculty of Science and Technology and MESA$^+$ Institute for Nanotechnology, University of Twente, P.O. Box 217, 7500 AE Enschede, The Netherlands}
\author{Anton A. Starikov}
\affiliation{Faculty of Science and Technology and MESA$^+$ Institute for Nanotechnology, University of Twente, P.O. Box 217, 7500 AE Enschede, The Netherlands}
\author{Arne Brataas}
\affiliation{Department of Physics, Norwegian University of Science and Technology, NO-7491 Trondheim, Norway}
\author{Paul J. Kelly}
\affiliation{Faculty of Science and Technology and MESA$^+$ Institute for Nanotechnology, University of Twente, P.O. Box 217, 7500 AE Enschede, The Netherlands}
\begin{abstract}
The precession and damping of a collinear magnetization displaced from its equilibrium are well described by the Landau-Lifshitz-Gilbert equation. The theoretical and experimental complexity of noncollinear magnetizations is such that it is not known how the damping is modified by the noncollinearity. We use first-principles scattering theory to investigate transverse domain walls (DWs) of the important ferromagnetic alloy Ni$_{80}$Fe$_{20}$ and show that the damping depends not only on the magnetization texture but also on the specific dynamic modes of Bloch and N{\'e}el DWs in ways that were not theoretically predicted. Even in the highly disordered Ni$_{80}$Fe$_{20}$ alloy, the damping is found to be remarkably nonlocal. 
\end{abstract}
\pacs{
72.25.Rb,         
75.60.Ch,	       
75.78.-n,          
75.60.Jk           
}
\maketitle

{\color{red}\it Introduction.}---The key common ingredient in various proposed nanoscale spintronics devices involving magnetic droplet solitons~\cite{Mohseni:sc13}, skyrmions~\cite{Yu:nat10,Fert:natn13}, or magnetic domain walls (DWs)~\cite{Thomas:sc10,Franken:natn12}, is a noncollinear magnetization that can be manipulated using current-induced torques (CITs)~\cite{Brataas:natm12}. Different microscopic mechanisms have been proposed for the CIT including spin transfer~\cite{Slonczewski:jmmm96,Berger:prb96}, spin-orbit interaction with broken inversion symmetry in the bulk or at interfaces~\cite{Miron:natm10,Emori:natm13,Boulle:prl13}, the spin-Hall effect~\cite{Liu:sc12} or proximity-induced anisotropic magnetic properties in adjacent normal metals~\cite{Ryu:natn13}. Their contributions are hotly debated but can only be disentangled if the Gilbert damping torque is accurately known. This is not the case \cite{Nembach:prl13}. Theoretical work  \cite{Foros:prb08,Tserkovnyak:prb09a,Zhang:prl09,Hankiewicz:prb08,Tserkovnyak:prb09b} suggesting that noncollinearity can modify the Gilbert damping due to the absorption of the pumped spin current by the adjacent precessing magnetization has stimulated experimental efforts to confirm this quantitatively~\cite{Nembach:prl13,Li:arxiv14}. In this Letter, we use first-principles scattering calculations to show that the Gilbert damping in a noncollinear alloy can be significantly enhanced depending on the particular precession modes and surprisingly, that even in a highly disordered alloy like Ni$_{80}$Fe$_{20}$, the nonlocal character of the damping is very substantial. Our findings are important for understanding field- and/or current-driven noncollinear magnetization dynamics and for designing new spintronics devices.

\begin{figure}[!b]
\begin{center}
\includegraphics[width=0.9\columnwidth]{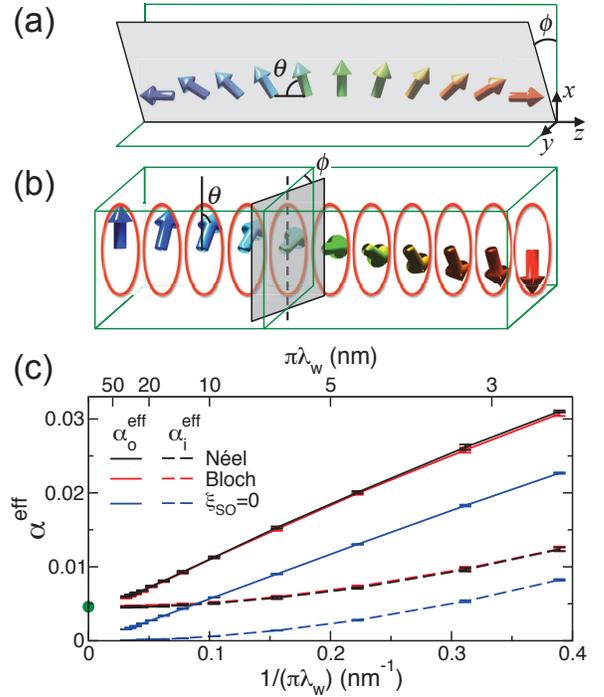}
\caption{(color online). Sketch of N{\'e}el (a) and Bloch (b) DWs. (c) Calculated effective Gilbert damping parameters for Permalloy DWs (N{\'e}el, black lines; Bloch, red lines) as a function of the inverse of the DW width $\lambda_w$. Without spin-orbit coupling, calculations for the two DW types yield the same results (blue lines). The green dot represents the value of Gilbert damping calculated for collinear Permalloy. For each value of $\lambda_w$, we typically consider 8 different disorder configurations and the error bars are a measure of the spread of the results.
}
\label{fig:sketch}
\end{center}
\end{figure}

{\color{red}\it Gilbert damping in Ni$_{80}$Fe$_{20}$ DWs.}---Gilbert damping is in general described by a symmetric $3\times3$ tensor. For a substitutional, cubic binary alloy like Permalloy, Ni$_{80}$Fe$_{20}$, this tensor is essentially diagonal and isotropic and reduces to  scalar form when the magnetization is collinear. A value of this dimensionless scalar calculated from first-principles, $\alpha_{\rm coll}=0.0046$, is in good agreement with values extracted from room temperature experiments that range between 0.004 and 0.009~\cite{Starikov:prl10}. In a one-dimensional (1D) transverse DW, the Gilbert damping tensor is still diagonal but, as a consequence of the lowered symmetry~\cite{Hals:prb14}, it contains two unequal components. The magnetization in static N{\'e}el or Bloch DWs lies inside well defined planes that are illustrated in Fig.~\ref{fig:sketch}. An angle $\theta$ represents the in-plane rotation with respect to the magnetization in the left domain and it varies from 0 to $\pi$ through a 180$^{\circ}$ DW. If the plane changes in time, as it does when the magnetization precesses, an angle $\phi$ can be used to describe its rotation. We define an out-of-plane damping component $\alpha_{\rm o}$ corresponding to variation in $\phi$, and an in-plane component $\alpha_{\rm i}$ corresponding to time-dependent $\theta$. Rigid translation of the DW, i.e. making the DW center $r_w$ vary in time, is a specific example of the latter. 

For Walker-profile DWs~\cite{Yuan:prl12}, an effective (dimensionless) in-plane ($\alpha_{\rm i}^{\rm eff}$) and out-of-plane damping ($\alpha_{\rm o}^{\rm eff}$) can be calculated in terms of the scattering matrix $\mathbf S$ of the system using the scattering theory of magnetization dissipation~\cite{Brataas:prl08,*Hals:prl09,*Brataas:prb11,SM}. Both calculated values are plotted in Fig.~\ref{fig:sketch}(c) as a function of the inverse DW width $1/\lambda_w$ for N{\'e}el and Bloch DWs. Results with the spin-orbit coupling (SOC) artificially switched off are shown for comparison; because spin space is then decoupled from real space, the results for the two DW profiles are identical and both $\alpha_{\rm i}^{\rm eff}$ and $\alpha_{\rm o}^{\rm eff}$ vanish in the large $\lambda_w$ limit confirming that SOC is the origin of intrinsic Gilbert damping for collinear magnetization. With SOC switched on, N{\'e}el and Bloch DWs have identical values within the numerical accuracy, reflecting the negligibly small magnetocrystalline anisotropy in Permalloy. Both $\alpha_{\rm i}^{\rm eff}$ and $\alpha_{\rm o}^{\rm eff}$ approach the collinear value $\alpha_{\rm coll}$~\cite{Starikov:prl10}, shown as a green dot in the figure, in the wide DW limit. For finite widths, they exhibit a quadratic and a predominantly linear dependence on $1/(\pi\lambda_w)$, respectively, both with and without SOC; for large values of $\lambda_w$, there is a hint of nonlinearity in $\alpha_{\rm o}^{\rm eff}(\lambda_w)$. However, phenomenological theories~\cite{Foros:prb08,Tserkovnyak:prb09a,Zhang:prl09} predict that $\alpha_{\rm i}^{\rm eff}$ should be independent of $\lambda_w$ and equal to $\alpha_{\rm coll}$ while $\alpha_{\rm o}^{\rm eff}$ should be a quadratic function of the magnetization gradient. Neither of these predicted behaviours is observed in Fig.~\ref{fig:sketch}(c) indicating that existing theoretical models of texture-enhanced Gilbert damping need to be reexamined.

The $\alpha^{\rm eff}$ shown in Fig.~\ref{fig:sketch}(c) is an effective damping constant because the magnetization gradient $d\theta/dz$ of a Walker profile DW is inhomogeneous. Our aim in the following is to understand the physical mechanisms of texture-enhanced Gilbert damping with a view to determining how the local damping depends on the magnetization gradient, as well as the corresponding parameters for Permalloy, and finally expressing these in a form suitable for use in micromagnetic simulations.

\begin{figure}[tbp]
\begin{center}
\includegraphics[width=\columnwidth]{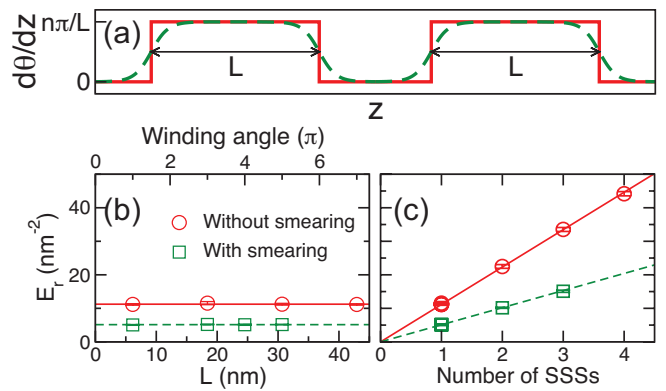}
\caption{(color online). (a) Sketch of the magnetization gradient for two SSSs separated by collinear magnetization with (green, dashed) and without (red, solid) a broadening of the magnetization gradient at the ends of the SSSs. The length of each segment is $L$. (b) Calculated energy pumping $E_r$ as a function of $L$ for a single Permalloy Bloch-DW-type SSS without SOC. The upper horizontal axis shows the total winding angle of the SSS. (c) Calculated energy pumping $E_r$ without SOC as a function of the number of SSSs that are separated by a stretch of collinear magnetization.}
\label{fig:inplane}
\end{center}
\end{figure}

{\color{red}\it In-plane damping $\alpha_{\rm i}$.}---To get a clearer picture of how the in-plane damping depends on the gradient, we calculate the energy pumping $E_r\equiv\mathrm{Tr}\left(\frac{\partial\mathbf S}{\partial r_s}\frac{\partial\mathbf S^{\dagger}}{\partial r_s}\right)$ for a finite length $L$ of a Bloch-DW-type spin spiral (SS) centered at $r_s$. In this SS segment (SSS), $d\theta/dz$ is constant except at the ends. Figure~\ref{fig:inplane}(b) shows the results calculated without SOC for a single Permalloy SSS with $d\theta/dz=6^{\circ}$ per atomic layer; Fig.~\ref{fig:sketch}(c) shows that SOC does not influence the quadratic behaviour essentially. $E_r$ is  seen to be independent of $L$ indicating there is no dissipation when $d\theta/dz$ is constant in the absence of SOC. In this case, the only contribution arises from the ends of the SSS where $d\theta/dz$ changes abruptly; see Fig.~\ref{fig:inplane}(a). If we replace the step function of $d\theta/dz$ by a Fermi-like function with a smearing width equal to one atomic layer, $E_r$ decreases significantly (green squares). For multiple SSSs separated by collinear magnetization, we find that $E_r$ is proportional to the number of segments; see Fig.~\ref{fig:inplane}(c). 

What remains is to understand the physical origin of the damping at the ends of the SSSs. Rigid translation of a SSS or of a DW allows for a dissipative spin current $\mathbf j_s''\sim-\mathbf m\times\partial_z\partial_t\mathbf m$ that breaks time-reversal symmetry~\cite{Tserkovnyak:prb09b}. The divergence of $\mathbf j_s''$ gives rise to a local dissipative torque, whose transverse component is the enhancement of the  in-plane Gilbert damping from the magnetization texture. After straightforward algebra, we obtain the texture-enhanced in-plane damping torque
\begin{eqnarray}
\alpha''\left[\left(\mathbf m\cdot\partial_z\partial_t\mathbf m\right)\mathbf m\times\partial_z\mathbf m-\mathbf m\times\partial^2_z\partial_t\mathbf m\right],\label{eq:inplane}
\end{eqnarray}
where $\alpha''$ is a material parameter with dimensions of length squared. In 1D SSs or DWs, Eq.~(\ref{eq:inplane}) leads to the local energy dissipation rate $\dot E(\mathbf r)=(\alpha''M_s/\gamma)\partial_t\theta\partial_t(d^2\theta/dz^2)$~\cite{SM}, where $M_s$ is the saturation magnetization and $\gamma=g\mu_B/\hbar$ is the gyromagnetic ratio expressed in terms of the Land{\'e} $g$-factor and the Bohr magneton $\mu_B$. This results shows explicitly that the in-plane damping enhancement is related to finite $d^2\theta/dz^2$. Using the calculated data in Fig.~\ref{fig:sketch}(c), we extract a value for the coefficient $\alpha''=0.016$~nm$^2$ that is independent of specific textures $\mathbf m(\mathbf r)$~\cite{SM}.

\begin{figure}[t]
\begin{center}
\includegraphics[width=0.85\columnwidth]{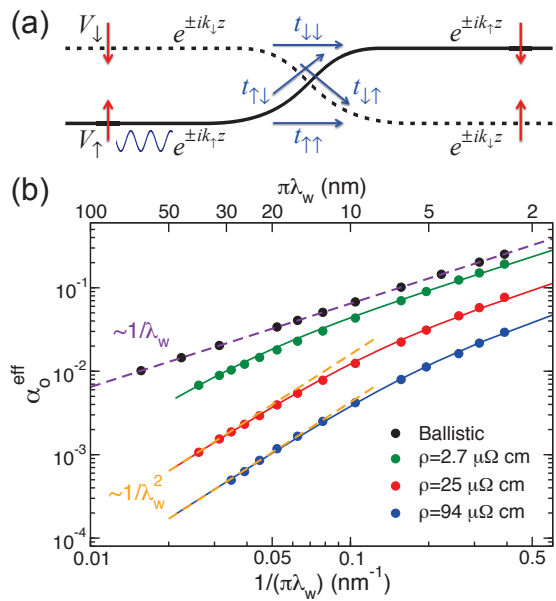}
\caption{(color online). (a) Cartoon of electronic transport in a two-band, free-electron DW. The global quantization axis of the system is defined by the majority and minority spin states in the left domain. (b) Calculated $\alpha_{\rm o}^{\rm eff}$ for two-band free-electron DWs as a function of $1/(\pi\lambda_w)$ on a log-log scale. The black circles show the calculated results for the clean DWs, which are in perfect agreement with the analytical model Eq.~(\ref{eq:free}), shown as a dashed violet line. When disorder (characterized by the resistivity $\rho$ calculated for the corresponding collinear magnetization) is introduced, $\alpha_{\rm o}^{\rm eff}$ shows a transition from a linear dependence on $1/\lambda_w$ for narrow DWs to a quadratic behaviour for wide DWs. The solid lines are fits using Eq.~(\ref{eq:fit}). The dashed orange lines illustrate quadratic behaviour. 
}
\label{fig:free}
\end{center}
\end{figure}

{\color{red}\it Out-of-plane damping $\alpha_{\rm o}$.}---We begin our analysis of the out-of-plane damping with a simple two-band free-electron DW model~\cite{SM}. Because the linearity of the damping enhancement does not depend on SOC, we examine the SOC free case for which there is no difference between N{\'e}el and Bloch DW profiles and we use N{\'e}el DWs in the following. Without disorder, we can use the known $\phi$-dependence of the scattering matrix for this model \cite{Duine:prb09} to obtain $\alpha_{\rm o}^{\rm eff}$ analytically,
\begin{eqnarray}
\alpha_{\rm o}^{\rm eff}&=&\frac{g\mu_B}{4\pi AM_s\lambda_w}\sum_{\mathbf k_\|}\left(\left\vert r_{\uparrow\downarrow}^{\mathbf k_\|}\right\vert^2+\left\vert r_{\downarrow\uparrow}^{\mathbf k_\|}\right\vert^2+\left\vert t_{\uparrow\downarrow}^{\mathbf k_\|}\right\vert^2+\left\vert t_{\downarrow\uparrow}^{\mathbf k_\|}\right\vert^2\right)\nonumber\\&\approx&\frac{g\mu_B}{4\pi AM_s\lambda_w}\frac{h}{e^2}G_{\rm Sh},\label{eq:free}.
\end{eqnarray}
where $A$ is the cross sectional area and the convention used for the reflection ($r$) and transmission ($t$) probability amplitudes is shown in Fig.~\ref{fig:free}(a). Note that $\vert t_{\uparrow\downarrow}^{\mathbf k_\|}\vert^2$ and $\vert t_{\downarrow\uparrow}^{\mathbf k_\|}\vert^2$ are of the order of unity and much larger than the other two terms between the brackets unless the exchange splitting is very large and the DW width very small. It is then a good approximation to replace the quantities in brackets by the number of propagating modes at $\mathbf k_\|$ to obtain the second line of Eq.~(\ref{eq:free}), where $G_{\rm Sh}$ is the Sharvin conductance that only depends on the free-electron density. Equation~(\ref{eq:free}) shows analytically that $\alpha_{\rm o}^{\rm eff}$ is proportional to $1/\lambda_w$ in the ballistic regime. This is reproduced by the results of numerical calculations for ideal free-electron DWs shown as black circles in Fig.~\ref{fig:free}(b). 

Introducing site disorder~\cite{Nguyen:prl08} into the free-electron model results in a finite resistivity. The out-of-plane damping calculated for disordered free-electron DWs exhibits a transition as a function of its width. For narrow DWs (ballistic limit), $\alpha_{\rm o}^{\rm eff}$ is inversely proportional to $\lambda_w$ and the green, red and blue circles in Fig.~\ref{fig:free}(b) tend to become parallel to the violet line for small values of $\lambda_w$. If $\lambda_w$ is sufficiently large, $\alpha_{\rm o}^{\rm eff}$ becomes proportional to $\lambda_w^{-2}$ in agreement with phenomenological predictions \cite{Foros:prb08,Tserkovnyak:prb09a,Zhang:prl09} where the diffusive limit is assumed. This demonstrates the different behaviour of $\alpha_{\rm o}^{\rm eff}$ in these two regimes.

We can construct an expression that describes both the ballistic and diffusive regimes by introducing an explicit spatial correlation in the nonlocal form of the out-of-plane Gilbert damping tensor that was derived using the fluctuation-dissipation theorem \cite{Foros:prb08}
\begin{eqnarray}
&&\left[\alpha_{\rm o}\right]_{ij}(\mathbf r,\mathbf r')=\alpha_{\rm coll}\delta_{ij}\delta(\mathbf r-\mathbf r')+\alpha'D(\mathbf r,\mathbf r';l_0)  \nonumber\\
&&~~~~~~~~\times\left[\mathbf m(\mathbf r)\times\partial_z\mathbf m(\mathbf r)\right]_i\left[\mathbf m(\mathbf r')\times\partial_{z'}\mathbf m(\mathbf r')\right]_j.\label{eq:fit}
\end{eqnarray}
Here $\alpha'$ is a material parameter with dimensions of length squared and $D$ is a correlation function with an effective correlation length $l_0$. In practice, we use $D(\mathbf r,\mathbf r';l_0)=\frac{1}{\sqrt{\pi}Al_0}e^{-(z-z')^2/l_0^2}$, which reduces to $\delta(\mathbf r-\mathbf r')$ in the diffusive limit ($l_0\ll\lambda_w$) and reproduces earlier results \cite{Foros:prb08,Tserkovnyak:prb09a,Zhang:prl09}. In the ballistic limit, both $\alpha'$ and $l_0$ are infinite, but the product $\alpha'D(\mathbf r,\mathbf r';l_0)=\alpha'/(\sqrt{\pi}Al_0)$ is finite and related to the Sharvin conductance of the system~\cite{Note4}, consistent with Eq.~(\ref{eq:free}). We then fit the calculated values of $\alpha_{\rm o}^{\rm eff}$ shown in Fig.~\ref{fig:free}(b) using Eq.~(\ref{eq:fit})~\cite{SM}.
With the parameters $\alpha'$ and $l_0$ listed in Table~\ref{tab:parameters}, the fit is seen to be excellent over the whole range of $\lambda_w$.
\begin{table}[b]
\caption{Fit parameters used to describe the damping shown in Fig.~\ref{fig:sketch} for Permalloy DWs and in Fig.~\ref{fig:free} for free-electron DWs with Eq.~\eqref{eq:fit}. The resistivity is determined for the corresponding collinear magnetization.
}
\begin{ruledtabular}
\begin{tabular}{cddd}
System & \multicolumn{1}{c}{$\rho$ ($\mu\Omega$ cm) } &
                      \multicolumn{1}{c}{$\alpha'$ (nm$^2$)} &
                                     \multicolumn{1}{c}{$l_0$ (nm)}  \\
\hline
Free electron & 2.69 & 45.0 &13.8\\
Free electron & 24.8   & 1.96 & 4.50\\
Free electron & 94.3  & 0.324 & 2.78\\
\hline
Py ($\xi_{\rm SO}=0$) & 0.504  & 23.1 & 28.3 \\
Py ($\xi_{\rm SO}\ne0$) & 3.45   &  5.91 &13.1\\
\end{tabular}
\end{ruledtabular}
\label{tab:parameters}
\end{table}%
The out-of-plane damping enhancement arises from the pumped spin current $\mathbf j_s'\sim \partial_t\mathbf m\times\partial_z\mathbf m$ in a magnetization texture~\cite{Foros:prb08,Zhang:prl09}, where the magnitude of $\mathbf j'_s$ is related to the conductivity~\cite{Foros:prb08}. This is the reason why $\alpha'$ is larger in a system with a lower resistivity in Table~\ref{tab:parameters}. $l_0$ is a measure of how far the pumped transverse spin current can propagate before being absorbed by the local magnetization. It is worth distinguishing the relevant characteristic lengths in microscopic spin transport that define the diffusive regimes for different transport processes. The mean free path $l_m$ is the length scale for diffusive charge transport. The spin-flip diffusion length $l_{\rm sf}$ characterizes the length scale for diffusive transport of a longitudinal spin current, and $l_0$ is the corresponding length scale for transverse spin currents. Only when the system size is larger than the corresponding characteristic length can transport be described in a local approximation. 

We can use Eq.~(\ref{eq:fit}) to fit the calculated $\alpha_{\rm o}^{\rm eff}$ shown in Fig.~\ref{fig:sketch} for Permalloy DWs. The results are shown in Fig.~\ref{fig:py}. Since the values of $\alpha_{\rm o}^{\rm eff}$ we calculate for N{\'e}el and Bloch DWs are nearly identical, we take their average for the SOC case. Intuitively, we would expect the out-of-plane damping for a highly disordered alloy like Permalloy to be in the diffusive regime corresponding to a short $l_0$. But the fitted values of $l_0$ are remarkably large, as long as 28.3~nm without SOC. With SOC, $l_0$ is reduced to 13.1~nm implying that nonlocal damping can play an important role in nanoscale magnetization textures in Permalloy, whose length scale in experiment is usually about 100~nm and can be reduced to be even smaller than $l_0$ by manipulating the shape anisotropy of experimental samples~\cite{Boulle:mser11,BenHamida:epl11}. 

As shown in Table~\ref{tab:parameters}, $l_0$ is positively correlated with the conductivity. The large value of $l_0$ and the low resistivity of Permalloy can be qualitatively understood in terms of its electronic structure and spin-dependent scattering. The Ni and Fe potentials seen by majority-spin electrons around the Fermi level in Permalloy are almost identical \cite{SM} so that they are only very weakly scattered. The Ni and Fe potentials seen by minority-spin electrons are however quite different leading to strong scattering in transport. The strong asymmetric spin-dependent scattering can also be seen in the resistivity of Permalloy calculated without SOC, where $\rho_{\downarrow}/\rho_{\uparrow}>200$~\cite{Banhart:prb97,Starikov:prl10}. As a result, conduction in Permalloy is dominated by the weakly scattered majority-spin electrons resulting in a low total resistivity and a large value of $l_0$. This short-circuit effect is only slightly reduced by SOC-induced spin-flip scattering because the SOC in 3$d$ transition metals is in energy terms small compared to the bandwidth and exchange splitting. Indeed, $\alpha_{\rm o}^{\rm eff}-\alpha_{\rm coll}$ calculated with SOC (the red curve in Fig.~\ref{fig:py}) shows a greater curvature at large widths than without SOC, but is still quite different from the quadratic function characteristic of diffusive behaviour for the widest DWs we could study.

\begin{figure}
\begin{center}
\includegraphics[width=\columnwidth]{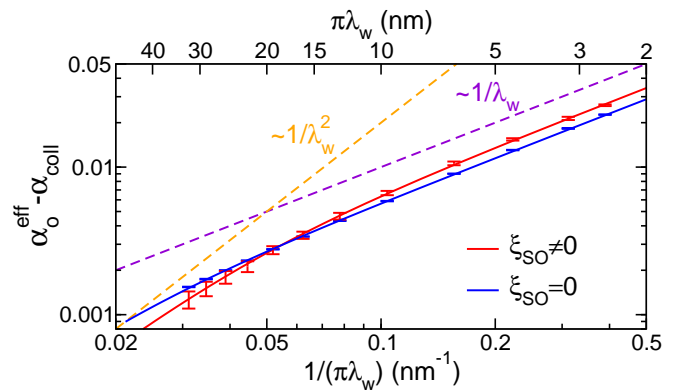}
\caption{(color online). Calculated out-of-plane damping $\alpha_{\rm o}^{\rm eff}-\alpha_{\rm coll}$ from Fig.~\ref{fig:sketch} plotted as a function of $1/(\pi\lambda_w)$ on a log-log scale. The solid lines are fitted using Eq.~(\ref{eq:fit}). The dashed violet and orange lines illustrate linear and quadratic behaviour, respectively. }
\label{fig:py}
\end{center}
\end{figure}

Both $\alpha_{\rm i}^{\rm eff}$ and $\alpha_{\rm o}^{\rm eff}$ originate from locally pumped spin currents proportional to $\mathbf m \times \partial_t\mathbf m$. Because of the spatially varying magnetization, the spin currents pumped to the left and right do not cancel exactly and the net spin current contains two components, $\mathbf j''_s\sim-\mathbf m\times\partial_z\partial_t\mathbf m$ \cite{Tserkovnyak:prb09b} and $\mathbf j'_s\sim\partial_t\mathbf m\times\partial_z\mathbf m$ \cite{Foros:prb08,Zhang:prl09}. 
For out-of-plane damping, $\partial_z\mathbf m$ is perpendicular to $\partial_t\mathbf m$ so there is large enhancement due to the lowest order derivative. For the rigid motion of a 1D DW, $\partial_z\mathbf m$ is parallel to $\partial_t\mathbf m$ so that $\mathbf j'_s$ vanishes. The enhancement of in-plane damping arising from $\mathbf j''_s$ due to the higher-order spatial derivative of magnetization is then smaller.

{\color{red}Conclusions.---}We have discovered an anisotropic texture-enhanced Gilbert damping in Permalloy DWs using first-principles calculations. The findings are expressed in a form [Eqs.~(\ref{eq:inplane}) and (\ref{eq:fit})] suitable for application to micromagnetic simulations of the dynamics of magnetization textures. The nonlocal character of the magnetization dissipation suggests that field and/or current driven DW motion, which is always assumed to be in the diffusive limit, needs to be reexamined. The more accurate form of the damping that we propose can be used to deduce the CITs in magnetization textures where the usual way to study them quantitatively is by comparing experimental observations with simulations. 

Current-driven DWs move with velocities that are proportional to $\beta/\alpha$ where $\beta$ is the nonadiabatic spin transfer torque parameter. The order of magnitude spread in values of $\beta$ deduced for Permalloy from measurements of the velocities of vortex DWs~\cite{Hayashi:apl08,Lepadatu:prl09,Eltschka:prl10,Chauleau:prb14} may be a result of assuming that $\alpha$ is a scalar constant. Our predictions can be tested by reexamining these studies using the expressions for $\alpha$ given in this paper as input to micromagnetic calculations.
 
\begin{acknowledgments}
We would like to thank Geert Brocks and Taher Amlaki for useful discussions. This work was financially supported by the ``Nederlandse Organisatie voor Wetenschappelijk Onderzoek'' (NWO) through the research programme of ``Stichting voor Fundamenteel Onderzoek der Materie'' (FOM) and the  supercomputer facilities of NWO ``Exacte Wetenschappen (Physical Sciences)''. It was also partly supported by the Royal Netherlands Academy of Arts and Sciences (KNAW). A.B. acknowledges the Research Council of Norway, grant no. 216700.
\end{acknowledgments}

%

\pagebreak
\begin{widetext}
\clearpage
\begin{center}
\textbf{\large Supplementary Material for ``Gilbert damping in noncollinear ferromagnets''}

\vspace{3mm}
Zhe Yuan,$^{1,\ast}$ Kjetil M. D. Hals,$^{2,3}$ Yi Liu,$^1$ Anton A. Starikov,$^1$ Arne Brataas,$^{2}$ and Paul J. Kelly$^1$

\vspace{2mm}
$^1${\small\it Faculty of Science and Technology and MESA$^+$ Institute for Nanotechnology, \\University of Twente, P.O. Box 217, 7500 AE Enschede, The Netherlands}

$^2${\small\it Department of Physics, Norwegian University of Science and Technology, NO-7491 Trondheim, Norway}

$^3${\small\it Niels Bohr International Academy and the Center for Quantum Devices, \\Niels Bohr Institute, University of Copenhagen, 2100 Copenhagen, Denmark}

\end{center}
\end{widetext}
\setcounter{equation}{0}
\setcounter{figure}{0}
\setcounter{table}{0}
\setcounter{page}{1}
\makeatletter
\renewcommand{\theequation}{S\arabic{equation}}
\renewcommand{\thefigure}{S\arabic{figure}}
\renewcommand{\thetable}{S\Roman{figure}}
\renewcommand{\bibnumfmt}[1]{[S#1]}
\renewcommand{\citenumfont}[1]{S#1}

\section{I. Computational details.}
Taking the concrete example of Walker profile domain walls (DWs), the effective (dimensionless) in-plane and out-of-plane damping parameters can be expressed in terms of the scattering matrix $\mathbf S$ of the system as, respectively,
\begin{eqnarray}
\alpha_{\rm i}^{\rm eff}&=&\frac{g\mu_B\lambda_w}{8\pi AM_s}\mathrm{Tr}\left(\frac{\partial\mathbf S}{\partial r_w}\frac{\partial\mathbf S^{\dagger}}{\partial r_w}\right),\label{eq:alphai}\\
\alpha_{\rm o}^{\rm eff}&=&\frac{g\mu_B}{8\pi AM_s\lambda_w}\mathrm{Tr}\left(\frac{\partial\mathbf S}{\partial \phi}\frac{\partial\mathbf S^{\dagger}}{\partial \phi}\right),\label{eq:alphao}
\end{eqnarray}
using the scattering theory of magnetization dissipation \cite{Hals:prl09s,Brataas:prb11s}. Here $g$ is the Land{\'e} $g$-factor, $\mu_B$ is the Bohr magneton, $\lambda_w$ denotes the DW width, $A$ is the cross sectional area, and $M_s$ is the saturation magnetization.

It is interesting to compare the scheme for calculating the Gilbert damping of DWs using Eqs.~(\ref{eq:alphai}) and (\ref{eq:alphao}) \cite{Hals:prl09s,Brataas:prb11s} with that used for collinear magnetization~\cite{Brataas:prl08s,Starikov:prl10s}. Both of them are based upon the energy pumping theory~\cite{Brataas:prl08s,Brataas:prb11s}. To calculate the damping $\alpha_{\rm coll}$ for the collinear case, the magnetization is made to precess uniformly and the local energy dissipation is homogeneous throughout the ferromagnet. The total energy loss due to Gilbert damping is then proportional to the volume of the ferromagnetic material and the homogenous local damping $\alpha_{\rm coll}$ can be determined from the damping per unit volume. When the magnetization of a DW is made to change either by moving its center $r_w$ or varying its orientation $\phi$, this results in a relatively large precession at the center of the DW; the further from the center, the less the magnetization changes. The local contribution to the total energy dissipation of the DW is weighted by the magnitude of the magnetization precession when $r_w$ or $\phi$ varies. For a fixed DW width, the total damping is not proportional to the volume of the scattering region but converges to a constant once the scattering region is large compared to the DW. In practice, $\alpha_{\rm i}^{\rm eff}$ and $\alpha_{\rm o}^{\rm eff}$ calculated using Eqs.~(\ref{eq:alphai}) and (\ref{eq:alphao}) are well converged for a scattering region 10 times longer than $\lambda_w$. Effectively, $\alpha^{\rm eff}$ can be regarded as a weighted average of the (dimensionless) damping constant in the region of a DW. In the wide DW limit, $\alpha^{\rm eff}_{\rm i}$ and $\alpha^{\rm eff}_{\rm o}$ both approach $\alpha_{\rm coll}$ with spin-orbit coupling (SOC) and vanish in its absence. 

To evaluate the effective Gilbert damping of a DW using Eqs.~(\ref{eq:alphai}) and (\ref{eq:alphao}), we attached semiinfinite (copper) leads to a finite length of Ni$_{80}$Fe$_{20}$ alloy (Permalloy, Py) and rotated the local magnetization to make a $180^{\circ}$ DW using the Walker profile. Specifically, we used $\mathbf m=(\mathrm{sech}\frac{z-r_w}{\lambda_w},0,\tanh\frac{z-r_w}{\lambda_w})$ for N{\'e}el DWs and $\mathbf m=(-\tanh\frac{z-r_w}{\lambda_w},-\mathrm{sech}\frac{z-r_w}{\lambda_w},0)$ for Bloch DWs. The scattering properties of the disordered region were probed by studying how Bloch waves in the Cu leads incident from the left or right sides were transmitted and reflected \cite{Starikov:prl10s,Yuan:prl12s}. The scattering matrix was obtained using a first-principles ``wave-function matching'' scheme \cite{Xia:prb06s} implemented with tight-binding linearized muffin-tin orbitals (TB-LMTOs)~\cite{Andersen:prb86s}. SOC was included using a Pauli Hamiltonian. The calculations were rendered tractable by imposing periodic boundary conditions transverse to the transport direction. It turned out that good results could be achieved even when these so-called ``lateral supercells'' were quite modest in size. In practice, we used $5\times5$ lateral supercells and the longest DW we considered was more than 500 atomic monolayers thick. After embedding the DW between collinear Py and Cu leads, the largest scattering region contained 13300 atoms. For every DW width, we averaged over about 8 random disorder configurations. 

A potential profile for the scattering region was constructed within the framework of the local spin density approximation of density functional theory as follows. For a slab of collinear Py binary alloy sandwiched between Cu leads, atomic-sphere-approximation (ASA) potentials~\cite{Andersen:prb86s} were calculated self-consistently without SOC using a surface Green's function (SGF) method implemented~\cite{Turek:97s} with TB-LMTOs. Charge and spin densities for binary alloy $A$ and $B$ sites were calculated using the coherent potential approximation~\cite{Soven:pr67s} generalized to layer structures \cite{Turek:97s}. For the scattering matrix calculation, the resulting ASA potentials were assigned randomly to sites in the lateral supercells subject to maintenance of the appropriate concentration of the alloy \cite{Xia:prb06s} and SOC was included. The exchange potentials are rotated in spin space~\cite{Wang:prb08s} so that the local quantization axis for each atomic sphere follows the DW profile. The DW width is determined in reality by a competition between interatomic exchange interactions and magnetic anisotropy.  For a nanowire composed of a soft magnetic material like Py, the latter is dominated by the shape anisotropy that arises from long range magnetic dipole-dipole interactions and depends on the nanowire profile. Experimentally it can be tailored by changing the nanowire dimensions leading to the considerable spread of reported DW widths~\cite{Boulle:mser11s}. 
In electronic structure calculations, that do not contain magnetic dipole-dipole interactions, we simulate a change of demagnetization energy by varying the DW width. In this way we can study the dependence of Gilbert damping on the magnetization gradient by performing a series of calculations for DWs with different widths.

For the self-consistent SGF calculations (without SOC), the two-dimensional (2D) Brillouin zone (BZ) corresponding to the $1\times1$ interface unit cell was sampled with a $120\times120$ grid. The transport calculations including SOC were performed with a $32\times32$ 2D BZ grid for a $5\times5$ lateral supercell, which is equivalent to a $160\times160$ grid in the $1\times1$ 2D BZ.

\section{II. Extracting $\alpha''$}
We first briefly derive the form of the in-plane damping. It has been argued phenomenologically \cite{Tserkovnyak:prb09cs} that for a noncollinear magnetization texture varying slowly in time the lowest order term in an expansion of the transverse component of the spin current in spatial and time derivatives that breaks time-reversal symmetry and is therefore dissipative is 
\begin{eqnarray}
\mathbf j''_s=-\eta\mathbf m\times\partial_z\partial_t\mathbf m,
\end{eqnarray}
where $\eta$ is a coefficient depending on the material and $\mathbf m$ is a unit vector in the direction of the magnetization. The divergence of the spin current,
\begin{eqnarray}
\partial_z\mathbf j''_s=-\eta\left(\partial_z\mathbf m\times\partial_z\partial_t\mathbf m+\mathbf m\times\partial^2_z\partial_t\mathbf m\right), \label{eq:disstorque}
\end{eqnarray}
gives the corresponding dissipative torque exerted on the local magnetization.
While the second term in brackets in Eq.~(\ref{eq:disstorque}) is perpendicular to $\mathbf m$, the first term contains both perpendicular and parallel components. Since we are only interested in the transverse component of the torque, we subtract the parallel component to find the damping torque 
\begin{eqnarray}
\bm{\tau}''&=&-\eta\left\{
(1 - \mathbf m \mathbf m) \cdot\left( \partial_z\mathbf m\times\partial_z\partial_t\mathbf m \right) 
 +\mathbf m\times\partial^2_z\partial_t\mathbf m\right\}\nonumber\\
&=&-\eta\left\{\left[\mathbf m\times\left(\partial_z\mathbf m\times\partial_z\partial_t\mathbf m\right)\right]\times\mathbf m+\mathbf m\times\partial^2_z\partial_t\mathbf m\right\}\nonumber\\
&=&\eta\left[(\mathbf m\cdot\partial_z\partial_t\mathbf m)\mathbf m\times\partial_z\mathbf m-\mathbf m\times\partial^2_z\partial_t\mathbf m\right].
\end{eqnarray} 
The Landau-Lifshitz-Gilbert equation including the damping torque $\bm{\tau}''$ reads
\begin{eqnarray}
\partial_t\mathbf m&=&-\gamma\mathbf m\times\mathbf H_{\rm eff}+\alpha_{\rm coll}\mathbf m\times\partial_t\mathbf m+\frac{\gamma\bm{\tau}''}{M_s} \nonumber\\
&=&-\gamma\mathbf m\times\mathbf H_{\rm eff}+\alpha_{\rm coll}\mathbf m\times\partial_t\mathbf m\nonumber\\
&&+\alpha''\left[(\mathbf m\cdot\partial_z\partial_t\mathbf m)\mathbf m\times\partial_z\mathbf m-\mathbf m\times\partial^2_z\partial_t\mathbf m\right]\label{eq:LLG},
\end{eqnarray}
where the in-plane damping parameter $\alpha''\equiv\gamma\eta/M_s$ has the dimension of length squared.

In the following, we explain how $\alpha''$ can be extracted from calculations on Walker DWs and show that it is applicable to other profiles. The formulation is essentially independent of the DW type (Bloch or N{\'e}el) and we use a Bloch DW in the following derivation for which
\begin{equation}
\mathbf m(z)=\left[\cos\theta(z),\sin\theta(z),0\right],\label{eq:bloch}
\end{equation}
where $\theta(z)$ represents the in-plane rotation (see Fig.~1 in the paper). 
The local energy dissipation associated with a time-dependent $\theta$ is given by \cite{Brataas:prb11s}
\begin{eqnarray}
&&\frac{\gamma}{M_s}\dot{E}(z)=\alpha_{\rm coll}\partial_t\mathbf m\cdot\partial_t\mathbf m\nonumber\\
&&~+\alpha''\left[\left(\mathbf m\cdot\partial_z\partial_t\mathbf m\right)\partial_t\mathbf m\cdot\partial_z\mathbf m-\partial_t\mathbf m\cdot\partial^2_z\partial_t\mathbf m\right].
\end{eqnarray}
For the one-dimensional profile Eq.~(\ref{eq:bloch}), this can be simplified as
\begin{eqnarray}
&&\frac{\gamma}{M_s}\dot{E}(z)=\alpha_{\rm coll}\left(\frac{d\theta}{dt}\right)^2-\alpha''\frac{d\theta}{dt}\frac{d}{dt}\left(\frac{d^2{\theta}}{dz^2}\right).\label{eq:ipe}
\end{eqnarray}
Substituting into Eq.~(\ref{eq:ipe}) the Walker profile 
\begin{equation}
\theta(z)=-\frac{\pi}{2}-\arcsin\left(\tanh\frac{z-r_w}{\lambda_w}\right),
\end{equation}
that we used in the calculations, we obtain for the total energy dissipation associated with the motion of a rigid DW for which $\dot\theta=\dot r_w d\theta/d r_w$,
\begin{eqnarray}
\dot E=\int d^3 r\,\dot{E}(z)=\frac{2M_s A}{\gamma\lambda_w}\left(\alpha_{\rm coll}+\frac{\alpha''}{3\lambda_w^2}\right)\dot{r}_w^2.\label{eq:walkerdiss}
\end{eqnarray}
Comparing this to the energy dissipation expressed in terms of the effective in-plane damping $\alpha_{\rm i}^{\rm eff}$~\cite{Brataas:prb11s}
\begin{eqnarray}
\dot E=\frac{2M_sA}{\gamma\lambda_w}\alpha_{\rm i}^{\rm eff}\dot r_w^2,
\end{eqnarray}
we arrive at
\begin{eqnarray}
\alpha_{\rm i}^{\rm eff}(\lambda_w)=\alpha_{\rm coll}+\frac{\alpha''}{3\lambda_w^2}.\label{eq:fitwalker}
\end{eqnarray}
Using Eq.~(\ref{eq:fitwalker}), we perform a least squares linear fitting of $\alpha_{\rm i}^{\rm eff}$ as a function of $\lambda_w^{-2}$ to obtain $\alpha_{\rm coll}$ and $\alpha''$. The fitting is shown in Fig.~\ref{fig:walker} and the parameters are listed in Table~{\ref{tab:walker}. Note that $\alpha_{\rm coll}$ is in perfect agreement with independent calculations for collinear Py \cite{Starikov:prl10s}.

\begin{figure}[t]
\begin{center}
\includegraphics[width=\columnwidth]{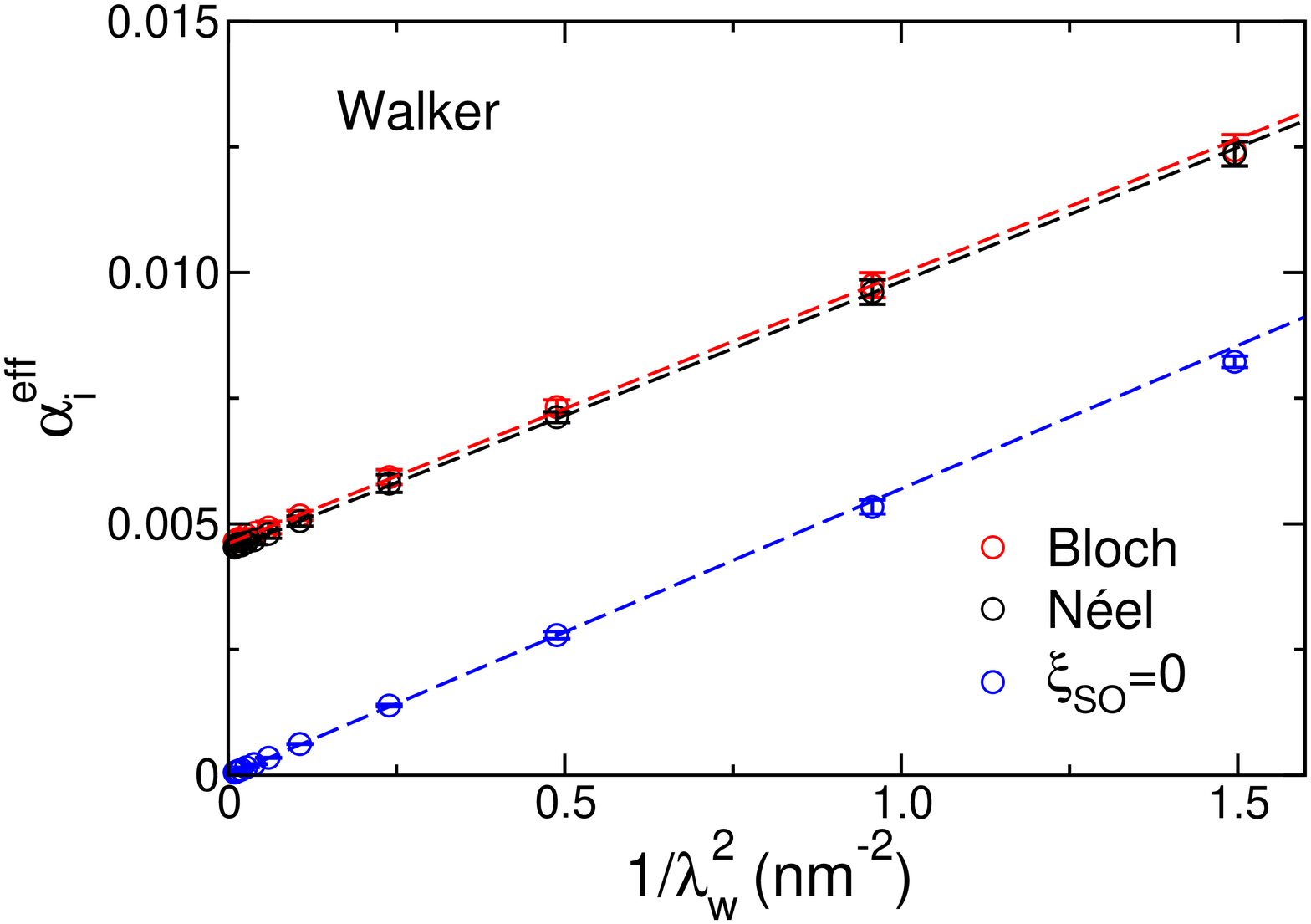}
\caption{Calculated $\alpha_{\rm i}^{\rm eff}$ for Walker-profile Permalloy DWs. N{\'e}el DWs: black circles, Bloch DWs: red circles. Without SOC, calculations for the two DW types yield the same results (blue circles). The dashed lines are linear fits using Eq.~(\ref{eq:fitwalker}). }
\label{fig:walker}
\end{center}
\end{figure}

To confirm that $\alpha''$ is independent of texture, we consider another analytical DW profile in which the in-plane rotation is described by a Fermi-like function,
\begin{eqnarray}
\theta(z)=-\pi+\frac{\pi}{1+e^{\frac{z-r_F}{\lambda_F}}}.\label{eq:fermi}
\end{eqnarray}
Here $r_F$ and $\lambda_F$ denote the DW center and width, respectively. Substituting Eq.~(\ref{eq:fermi}) into Eq.~(\ref{eq:ipe}), we find the energy dissipation for ``Fermi'' DWs to be
\begin{eqnarray}
\dot E=\frac{\pi^2 M_s A}{6\gamma\lambda_F}\left(\alpha_{\rm coll}+\frac{\alpha''}{5\lambda_F^2}\right)\dot{r}_F^2,
\end{eqnarray}
which suggests the effective in-plane damping 
\begin{equation}
\alpha_{\rm i}^{\rm eff}(\lambda_F)=\alpha_{\rm coll}+\frac{\alpha''}{5\lambda_F^2}.
\label{eq:prdfermi}
\end{equation}
Eq.~\eqref{eq:prdfermi} is plotted as solid lines in Fig.~\ref{fig:fermi} with the values of $\alpha_{\rm coll}$ and $\alpha''$ taken from Table~\ref{tab:walker}.
}

\begin{table}[b]
\caption{Fit parameters to describe the in-plane Gilbert damping in Permalloy DWs.}
\begin{ruledtabular}
\begin{tabular}{ccc}
DW type &  $\alpha_{\rm coll}$ &
                         $\alpha''$ (nm$^2$) \\
\hline
Bloch                      & (4.6$\pm$0.1)$\times10^{-3}$ &0.016$\pm$0.001 \\
N{\'e}el                   & (4.5$\pm$0.1)$\times10^{-3}$ &0.016$\pm$0.001 \\
$\xi_{\rm SO}$=0 & (2.0$\pm$1.0)$\times10^{-6}$ &0.017$\pm$0.001\\
\end{tabular}
\end{ruledtabular}
\label{tab:walker}
\end{table}%

Since the energy pumping can be expressed in terms of the scattering matrix $\mathbf S$ as 
\begin{equation}
\dot E=\frac{\hbar}{4\pi}\mathrm{Tr}\left(\frac{\partial \mathbf S}{\partial t}\frac{\partial \mathbf S^{\dagger}}{\partial t}\right)=\frac{\hbar}{4\pi}\mathrm{Tr}\left(\frac{\partial \mathbf S}{\partial r_F}\frac{\partial \mathbf S^{\dagger}}{\partial r_F}\right)\dot{r}_F^2,
\end{equation}
we can calculate the effective in-plane damping for a Fermi DW from the $\mathbf S$ matrix to be
\begin{eqnarray}
\alpha_{\rm i}^{\rm eff}=\frac{3\hbar\gamma\lambda_F}{2\pi^3 M_s A}\mathrm{Tr}\left(\frac{\partial \mathbf S}{\partial r_F}\frac{\partial\mathbf S^{\dagger}}{\partial r_F}\right).\label{eq:alphafermi}
\end{eqnarray}
We plot the values of $\alpha_{\rm i}^{\rm eff}$ calculated using the derivative of the scattering matrix Eq.~(\ref{eq:alphafermi}) as circles in Fig.~\ref{fig:fermi}. The good agreement between the circles and the solid lines demonstrates the validity of the form of the in-plane damping torque in Eq.~(\ref{eq:LLG}) and that the parameter $\alpha''$ does not depend on a specific magnetization texture. 

\begin{figure}[t]
\begin{center}
\includegraphics[width=\columnwidth]{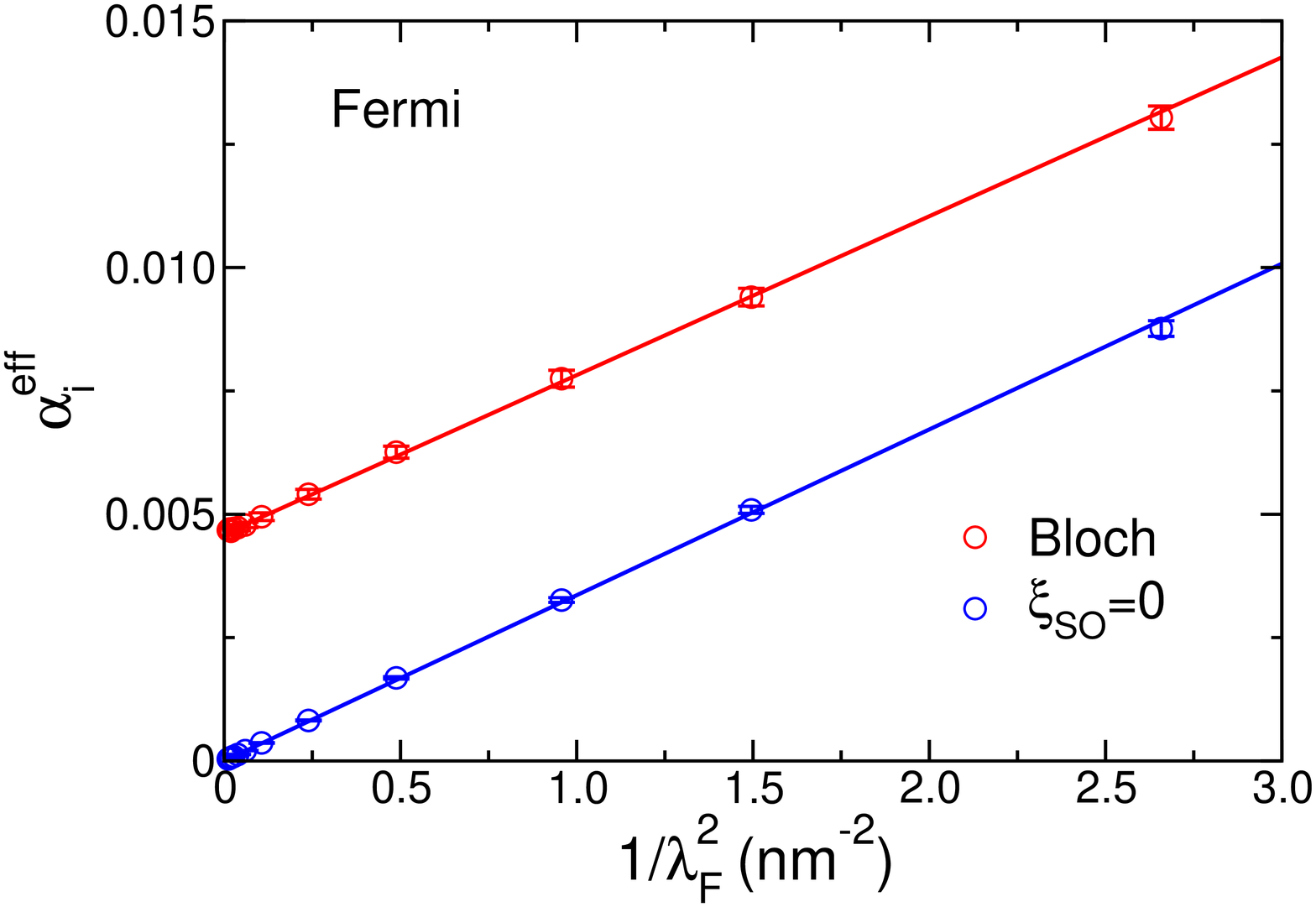}
\caption{Calculated $\alpha_{\rm i}^{\rm eff}$ for Permalloy Bloch DWs (red circles) with the Fermi profile Eq.~(\ref{eq:fermi}). The blue circles are results calculated without SOC. The solid lines are the analytical expression Eq.~\eqref{eq:prdfermi} using the parameters listed in Table~\ref{tab:walker}. }
\label{fig:fermi}
\end{center}
\end{figure}

\section{III. The free-electron model using muffin-tin orbitals}
\begin{figure}[b]
\centering
\includegraphics[width=\columnwidth]{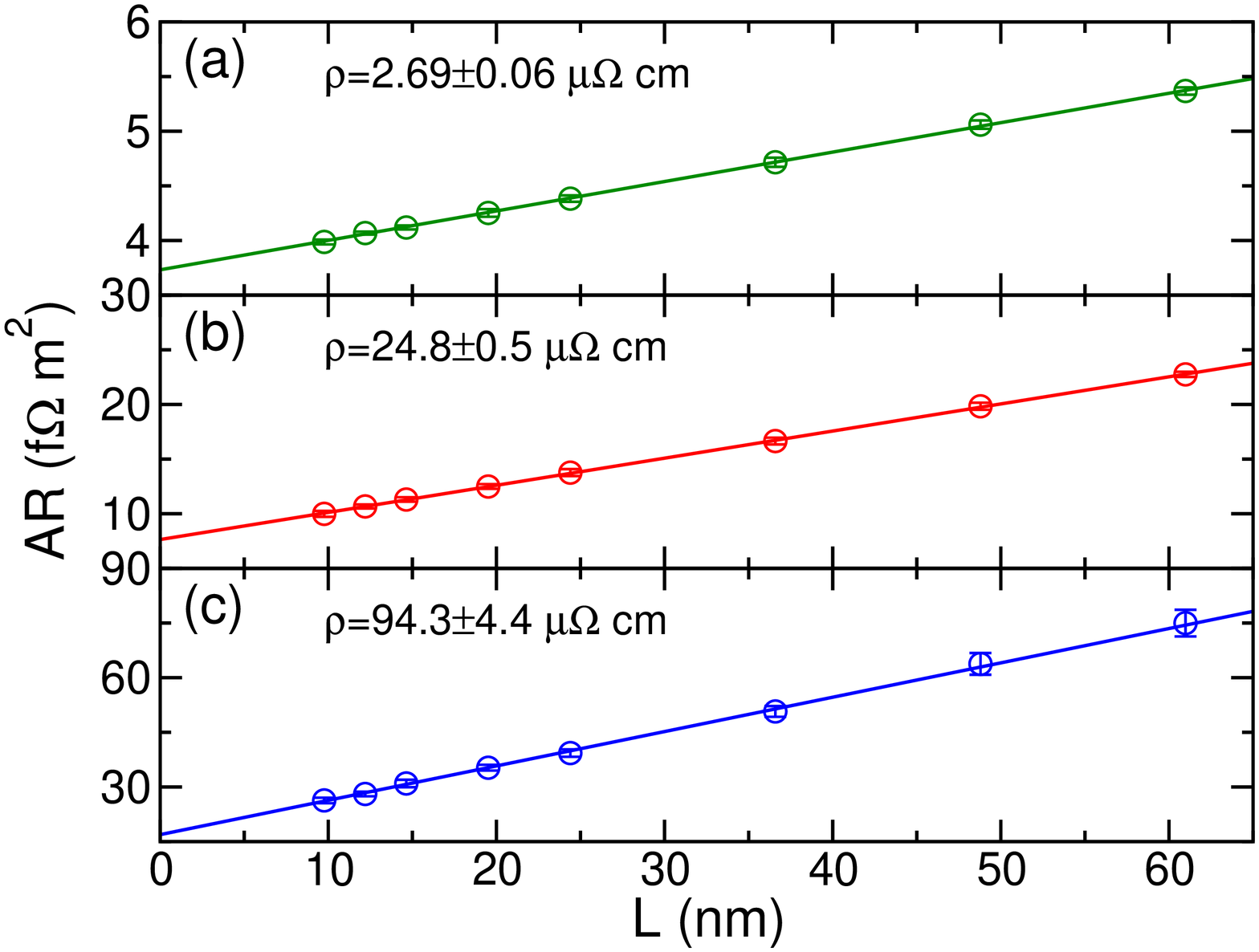}
\caption{Resistance calculated for the disordered free-electron model as a function of the length of the scattering region for three values of $V_0$, the disorder strength: 0.05 Ry (a), 0.15 Ry (b) and 0.25 Ry (c). The lines are the linear fitting used to determine the resistivity.} 
\label{fig:freerho}
\end{figure}
We take constant potentials, $V_{\uparrow}=-0.2$~Ry, $V_{\downarrow}=-0.1$~Ry inside atomic spheres with an exchange splitting $\Delta V =0.1$~Ry between majority and minority spins and a Fermi level $E_F=0$. The atomic spheres are placed on a face-centered cubic (fcc) lattice with the lattice constant of nickel, 3.52~{\AA}. The magnetic moment on each atom is then $0.072~\mu_B$. The transport direction is along the fcc [111]. In the scattering calculation, we use a 300$\times$300 $k$-point mesh in the 2D BZ. The calculated Sharvin conductances for majority and minority channels are 0.306 and 0.153 $e^2/h$ per unit cell, respectively, compared with analytical values of 0.305 and 0.153. 

To mimic disordered free-electron systems, we introduce a 5$\times$5 lateral supercell and distribute constant potentials uniformly in the energy range $[-V_0/2,V_0/2]$ where $V_0$ is some given strength~\cite{Nguyen:prl08s} and spatially at random on every atomic sphere in the scattering region. The calculated total resistance as a function of the length $L$ of the (disordered) scattering region is shown in Fig.~\ref{fig:freerho} with $V_0=0.05$~Ry (a), 0.15~Ry (b) and 0.25~Ry (c). The resistivity increases with the impurity strength as expected and can be extracted with a linear fitting $AR(L)=AR_0+\rho L$. For each system, we calculate about 10 random configurations and take the average of the calculated results. Well converged results are obtained using a $32\times32$ $k$-point mesh for the $5\times5$ supercell. 

\begin{widetext}
\section{IV. Fitting $\alpha'$ and $l_0$}
With a nonlocal Gilbert damping, $\bm{\alpha}(\mathbf r,\mathbf r')$, the energy dissipation rate is given by~\cite{Brataas:prb11s}
\begin{eqnarray}
\dot E=\frac{M_s}{\gamma}\int d^3r\,\dot{\mathbf m}(\mathbf r)\cdot\int d^3r'\,\bm{\alpha}(\mathbf r,\mathbf r')\cdot\dot{\mathbf m}(\mathbf r').\label{eq:edot}
\end{eqnarray}
If we consider the out-of-plane damping of a N{\'e}el DW, i.e. for which the angle $\phi$ varies in time (see Fig.~1 in the paper), we have
\begin{eqnarray}
\dot{\mathbf m}(\mathbf r)=\dot{\phi} \, \mathrm{sech}\frac{z-r_w}{\lambda_w}\hat y.\label{eq:mdot}
\end{eqnarray}
Considering again a Walker profile, we find the explicit form of the out-of-plane damping matrix element
\begin{eqnarray}
\alpha_{\rm o}(z,z')=\alpha_{\rm coll}\delta(z-z')+\frac{\alpha'}{\lambda_w^2} \mathrm{sech}\frac{z-r_w}{\lambda_w} \; \mathrm{sech}\frac{z'-r_w}{\lambda_w}\frac{1}{\sqrt{\pi}A l_0}e^{-(\frac{z-z'}{l_0})^2}.\label{eq:alphaoneel}
\end{eqnarray}
Substituting Eq.~(\ref{eq:alphaoneel}) and Eq.~(\ref{eq:mdot}) into Eq.~(\ref{eq:edot}), we obtain explicitly the energy dissipation rate
\begin{eqnarray}
\dot E=\frac{2M_sA\lambda_w}{\gamma}\alpha_{\rm coll}\dot{\phi}^2+\frac{M_sA\alpha'\dot{\phi}^2}{\sqrt{\pi}\gamma l_0\lambda_w^2}\int dz\,\mathrm{sech}^2\frac{z-r_w}{\lambda_w}\int dz'\,\mathrm{sech}^2\frac{z'-r_w}{\lambda_w}
e^{-(\frac{z-z'}{l_0})^2}.\label{eq:edot1}
\end{eqnarray}

The calculated effective out-of-plane Gilbert damping for a DW with the Walker profile is related to the energy dissipation rate as~\cite{Brataas:prb11s}
\begin{eqnarray}
\dot E=\frac{2M_sA\lambda_w}{\gamma}\alpha_{\rm o}^{\rm eff}\dot{\phi}^2.\label{eq:edot2}
\end{eqnarray}
Comparing Eqs.~(\ref{eq:edot1}) and (\ref{eq:edot2}), we arrive at
\begin{eqnarray}
\alpha_{\rm o}^{\rm eff}=\alpha_{\rm coll}+\frac{\alpha'}{2\sqrt{\pi}\lambda_w^3 l_0}\int dz\,\mathrm{sech}^2\frac{z-r_w}{\lambda_w}\int dz'\,\mathrm{sech}^2\frac{z'-r_w}{\lambda_w}e^{-(\frac{z-z'}{l_0})^2}.\label{eq:fit}
\end{eqnarray}
The last equation is used to fit $\alpha'$ and $l_0$ to $\alpha_{\rm o}^{\rm eff}$ calculated for different $\lambda_w$.  For Bloch DWs, it is straightforward to repeat the above derivation and find the same result, Eq.~(\ref{eq:fit}).
\end{widetext}

\section{V. Band structures of Ni and Fe in Permalloy}

\begin{figure}
\begin{center}
\includegraphics[width=\columnwidth]{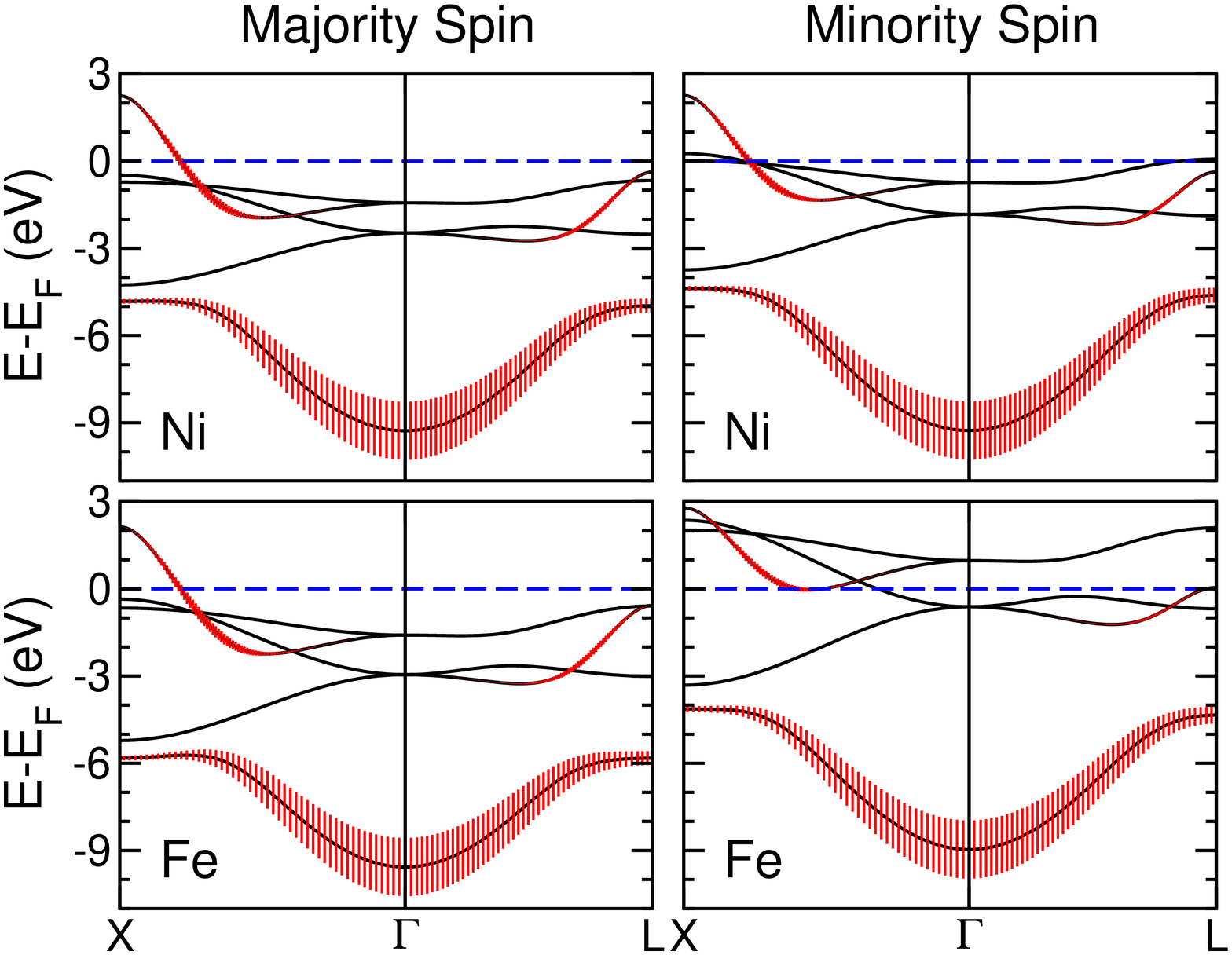}
\caption{Band structures calculated with the auxiliary Ni and Fe atomic sphere potentials and Fermi energy that were calculated self-consistently for Ni$_{80}$Fe$_{20}$ using the coherent potential approximation. The red bars indicates the amount of $s$ character in each band.}
\label{fig:py}
\vspace{-1em}
\end{center}
\vspace{-1em}
\end{figure}

In the coherent potential approximation (CPA) \cite{Soven:pr67s,Turek:97s}, the single-site approximation involves calculating auxiliary (spin-dependent) potentials for Ni and Fe self-consistently. In our transport calculations, these auxiliary potentials are distributed randomly in the scattering region. It is instructive to place the Ni potentials (for majority- and minority-spin electrons) on an fcc lattice and to calculate the band structure non-self-consistently. Then we do the same using the Fe potentials. The corresponding band structures are plotted in Fig.~\ref{fig:py}. At the Fermi level, where electron transport takes place, the majority-spin bands for Ni and Fe are almost identical, including their angular momentum character. This means that majority-spin electrons in a disordered alloy see essentially the same potentials on all lattice sites and are only very weakly scattered in transport by the randomly distributed Ni and Fe potentials. In contrast, the minority-spin bands are quite different for Ni and Fe. This can be understood in terms of the different exchange splitting between majority- and minority-spin bands; the calculated magnetic moments of Ni and Fe in Permalloy in the CPA are 0.63 and 2.61~$\mu_B$, respectively. The random distribution of Ni and Fe potentials in Permalloy then leads to strong scattering of minority-spin electrons in transport.

\end{document}